\documentclass[11pt,a4paper]{article}
\usepackage{comment}

\usepackage[margin=25mm]{geometry}
\usepackage[T1]{fontenc}
\usepackage{lmodern}

\usepackage{graphicx}
\usepackage{booktabs}
\usepackage{subcaption}

\usepackage{amsmath,amssymb}

\usepackage{authblk}
\usepackage{hyperref}
\usepackage[nameinlink,noabbrev]{cleveref}
\hypersetup{
  colorlinks=true,
  linkcolor=blue,
  urlcolor=blue,
  citecolor=blue
}

\usepackage{listings}
\usepackage{xcolor}
\usepackage{tikz}

\usepackage{orcidlink}

\usepackage{xurl}
\usepackage{placeins}
\usepackage{indentfirst}
\usepackage[
  backend=biber,
  style=numeric,
  sorting=none,
  maxbibnames=3,
  minbibnames=1,
  maxcitenames=2,
  mincitenames=1,
  doi=false,
  eprint=false,
]{biblatex}

\usepackage{xparse}
\NewDocumentCommand{\figprelim}{O{} O{3.2} m}{%
\begin{tikzpicture}
  \node[inner sep=0] (img) {\includegraphics[#1]{#3}};
  \node[rotate=30, scale=#2, text opacity=0.20] at (img.center)
    {\bfseries PRELIMINARY};
\end{tikzpicture}%
}

\makeatletter
\newcommand{\chan}[1]{%
  \begingroup
  \edef\x{\endgroup\noexpand\nolinkurl{\detokenize{#1}}}%
  \x
}
\makeatother

\title{
    Statistical-Uncertainty-Driven Selection of Evaluation
    Frequency for Time-Dependent Sensing Calibration:
    A Demonstration with KAGRA Data
}

\author[1,2]{Shingo Hido\orcidlink{0009-0009-0004-4170}
\thanks{E-mail: \texttt{shingo@icrr.u-tokyo.ac.jp}}
}
\author[2]{Takahiro Yamamoto\orcidlink{0000-0002-0808-4822}}
\author[3]{Dan Chen\orcidlink{0000-0003-1433-0716}}
\author[2]{Takahiro Sawada\orcidlink{0000-0001-5726-7150}}
\author[2]{Shinji Miyoki\orcidlink{0000-0002-1213-8416}}

\affil[1]{Department of Physics, The University of Tokyo, 7-3-1 Hongo, Bunkyo-ku, Tokyo 113-0033, Japan}
\affil[2]{Institute for Cosmic Ray Research, KAGRA Observatory, The University of Tokyo, 238 Higashi-Mozumi, Kamioka-cho, Hida City, Gifu 506-1205, Japan}
\affil[3]{Gravitational Wave Science Project, National Astronomical Observatory of Japan, 2-21-1 Osawa, Mitaka, Tokyo 181-8588, Japan}

\date{}

\begin{document}
\maketitle

\begin{abstract}
Accurate calibration of the gravitational-wave strain $h(t)$ is essential for both detection and astrophysical inference. In operating detectors, slow temporal variations in the sensing response are tracked using calibration lines, but practical constraints can prevent those lines from being injected at frequencies that are favorable for precise estimation of sensing-side parameters.
We present a statistical framework for preselecting evaluation frequencies under such constraints.
We apply this framework to KAGRA data from the first part of the fourth LIGO-Virgo-KAGRA Observing Run, for which the nominal cavity-pole frequency was about 18 Hz, while the sensing-side calibration line used in practice was injected at 32.7 Hz. For each candidate evaluation frequency, we construct the sensing function, quantify its segment-wise statistical uncertainty from empirical percentiles of the sample distribution, and rank the candidates using a score that combines the interval widths of the amplitude and phase.
When a 1\% amplitude interval width and a 1 degree phase interval width are weighted equally, 244 Hz is selected in all 4096 s analysis segments throughout the analyzed period.
Relative to the reference frequency of 32.7 Hz, the amplitude interval width is reduced to about one quarter over a broad frequency range, while the phase interval width remains broadly comparable.
We also assess the discrepancy introduced by frequency translation separately. These results suggest that the proposed method provides a useful statistical preselection framework for evaluation frequencies under practical operational constraints.

\end{abstract}

\clearpage

\FloatBarrier
\newpage
\section{Introduction}
\label{sec:intro}

Gravitational-wave detectors measure minute strains produced by astrophysical sources such as compact binary coalescences. To extract reliable astrophysical information from these signals, it is essential that the gravitational-wave strain $h(t)$ reconstructed from the detector output be accurately calibrated. Previous studies have shown that calibration bias can affect not only detections but also subsequent astrophysical inference, including parameter estimation and cosmological measurements~\cite{PhysRevD.85.064034,PhysRevD.111.063034,Hall_2019}.

KAGRA is a ground-based gravitational-wave detector featuring underground construction and cryogenic test masses~\cite{10.1093/ptep/ptaa125}, and is a member of the LIGO--Virgo--KAGRA network~\cite{Aasi_2015,Acernese_2015}. In this study, we analyze KAGRA data from the first part of the fourth LIGO-Virgo-KAGRA Observing Run (O4a)~\cite{Abac_2025}. Because the detector response varies over time during observations, calibration must track these variations to maintain accurate strain reconstruction. Calibration lines are used to monitor such temporal changes in the detector response through time-dependent correction factors (TDCFs)~\cite{Tuyenbayev_2017,Wade_2023}. On the sensing side, the quantities of interest are the relative optical gain $\kappa_C(t)$ and the cavity-pole frequency $f_{\mathrm{cp}}(t)$.

The choice of frequency used to evaluate these sensing-side quantities is particularly important for estimating $f_{\mathrm{cp}}(t)$. It is known that the precision of the cavity-pole estimate improves when the relevant calibration-line frequency is close to the nominal cavity-pole frequency~\cite{Tuyenbayev_2017}.
In KAGRA O4a, however, the nominal cavity-pole frequency was approximately 18~Hz because of the high finesse~\cite{10.1093/ptep/ptaa125}, whereas the sensing-side calibration line used in practice was placed at 32.7~Hz, since low-frequency injection with sufficient signal-to-noise ratio was not operationally feasible~\cite{LINE_info}.
This illustrates a practical situation in which the statistically favorable frequency for sensing-side estimation is not available for direct calibration-line injection.

Rather than changing the calibration-line injection frequency itself, we formulate the choice of evaluation frequency as a statistical preselection problem under such operational constraints. Using information obtained from the calibration line together with the reference model, we evaluate $\kappa_C(t)$ and $f_{\mathrm{cp}}(t)$ at candidate evaluation frequencies and compare the resulting statistical uncertainty of the sensing function. The aim of this study is to identify near-optimal evaluation frequencies under a statistical objective, without modifying the injected calibration-line frequency itself.

In the following, we first rank candidate evaluation frequencies by their statistical performance, and then assess separately the consistency of the translated results with the reference-frequency result. This separation allows the statistical optimization itself and the supplementary validation of translation-induced discrepancy to be discussed distinctly.


\section{Method}
\label{sec:method}

\subsection{Sensing-function model and TDCF evaluation at an arbitrary evaluation frequency}
\label{sec:sens}

The calibrated strain \(h(t)\) used in gravitational-wave analyses is constructed from the differential arm (DARM) displacement. Accurate reconstruction and evaluation of \(h(t)\) therefore require an appropriate description of the DARM control-loop response~\cite{Viets_2018}. During observation, externally induced DARM displacements are suppressed by the loop, and calibration reconstructs this displacement using signals within the loop together with a model of the loop response. Calibration lines are used to track temporal variations in the response and to estimate the TDCFs~\cite{Tuyenbayev_2017, Wade_2023}.

The DARM control loop is described by the sensing function \(C(f,t)\), the digital control filter \(D(f)\), and the actuation function \(A(f,t)\). In this study, we focus on the sensing function and model it in terms of the relative optical-gain variation \(\kappa_C(t)\) and the cavity-pole frequency \(f_{\mathrm{cp}}(t)\) as
\begin{equation}
\label{eq:C_st}
C(f,t)=\frac{\kappa_C(t)\,H_C}{1+i f/f_{\mathrm{cp}}(t)}\,F_C(f)e^{-2\pi i f\tau_C}.
\end{equation}
Here, \(H_C\) is the reference optical gain, \(F_C(f)\) represents the remaining frequency-dependent components, and \(\tau_C\) is the sensing delay~\cite{Viets_2018, 10.1093/ptep/ptab018}.
In the conventional TDCF estimation, \(\kappa_C(t)\) and \(f_{\mathrm{cp}}(t)\) are estimated at the calibration-line frequency using the complex response measured from the calibration line.

We extend the conventional formulation to an arbitrary evaluation frequency $f_e$ by translating the complex transfer function measured at the calibration-line frequency to fe using the reference-model ratio and then applying the same algebraic relations as in the conventional method~\cite{Tuyenbayev_2017, Wade_2023, Hido2026}. This allows the sensing-side quantities to be evaluated at an arbitrary frequency.

The sensing function constructed from the quantities evaluated at the evaluation frequency \(f_e\) is written as
\begin{equation}
\label{eq:C_fe_t}
C^{[f_e]}(f,t)
=
\frac{\kappa_C^{[f_e]}(t)\,H_C}{1 + i f / f_{\mathrm{cp}}^{[f_e]}(t)}
\,F_C(f)\,e^{-2\pi i f \tau_C},
\end{equation}
where the superscript \([f_e]\) indicates that the corresponding time series is obtained using the calibration quantities calculated at the evaluation frequency \(f_e\).
The choice of \(f_e\) therefore affects the resulting sensing function \(C^{[f_e]}(f,t)\), and thus the associated statistical uncertainty.
We next summarize the statistical spread of the samples of \(C^{[f_e]}(f,t)\) and define the score used to compare candidate evaluation frequencies.
This translation assumes consistency with the reference model. Possible model--measurement bias~\cite{Hido2026} and the resulting systematic discrepancy are assessed separately in Sec.~\ref{sec:supp}.

\subsection{Objective function and statistical uncertainty}
\label{sec:obj}

Using the sensing function \(C^{[f_e]}(f,t)\) defined in Eq.~(\ref{eq:C_fe_t}), we quantify the statistical uncertainty associated with each evaluation frequency \(f_e\) by summarizing the sample distribution of the sensing function for each candidate and defining a score to compare them.

For each analysis segment \(\mathcal{T}_j\), we evaluate the sensing function over times \(t \in \mathcal{T}_j\) and construct the sample set
\begin{equation}
\left\{
C^{[f_e]}(f,t)
\;\middle|\;
t \in \mathcal{T}_j
\right\}.
\label{eq:C_sample_set}
\end{equation}
We summarize the statistical uncertainty of the sensing function within each segment using empirical percentiles. For the magnitude, we define the relative width of the \(68\%\) interval as
\begin{equation}
W_{|C|}^{(j)}(f_e,f)
=
\frac{
Q_{84}\!\left(\left|C^{[f_e]}(f,t)\right|\right)
-
Q_{16}\!\left(\left|C^{[f_e]}(f,t)\right|\right)
}{
Q_{50}\!\left(\left|C^{[f_e]}(f,t)\right|\right)
},
\label{eq:Cmag_width}
\end{equation}
where \(Q_{16}\), \(Q_{50}\), and \(Q_{84}\) denote the 16th, 50th, and 84th percentiles of the time samples within the segment \(\mathcal{T}_j\), respectively. For the phase, we similarly define the \(68\%\) interval width as
\begin{equation}
W_{\arg C}^{(j)}(f_e,f)
=
Q_{84}\!\left(\arg C^{[f_e]}(f,t)\right)
-
Q_{16}\!\left(\arg C^{[f_e]}(f,t)\right).
\label{eq:Cphase_width}
\end{equation}
These quantify the \(68\%\) interval widths of the magnitude and phase distributions, respectively.

To summarize the quality of a given evaluation frequency \(f_e\) into a single score for each analysis segment \(\mathcal{T}_j\), we define
\begin{equation}
S^{(j)}(f_e,\sigma_{|A|},\sigma_\phi)
=
\frac{
\displaystyle
\int_{f_{\min}}^{f_{\max}}
w(f)
\left[
\frac{W_{|C|}^{(j)}(f_e,f)}{\sigma_{|A|}}
+
\frac{W_{\arg C}^{(j)}(f_e,f)}{\sigma_\phi}
\right]
\,d\log f
}{
\displaystyle
\int_{f_{\min}}^{f_{\max}} w(f)\,d\log f
}.
\label{eq:score_def}
\end{equation}
Here, \(w(f)\) is a frequency-dependent weight function, and \(\sigma_{|A|}\) and \(\sigma_{\phi}\) are normalization constants that control the relative contributions of the magnitude and phase terms, respectively. In this work, we adopt
\begin{equation}
w(f)=1,
\label{eq:weight_unity}
\end{equation}
so that the score is defined as a simple average over logarithmic frequency.
Multiplying \(\sigma_{|A|}\) and \(\sigma_{\phi}\) by the same factor changes only the overall scale of the score and does not affect the minimizing frequency.
Therefore, the preferred evaluation frequency depends only on the ratio \(\sigma_{\phi}/\sigma_{|A|}\).
In the main analysis, we adopt $\sigma_{\phi}/\sigma_{|A|}=1$ as a representative choice. This is motivated by the fact that previous $h(t)$ uncertainty estimates in gravitational-wave detectors often place the relative scale of phase uncertainty in degrees and amplitude uncertainty in percent near unity, although the exact ratio is frequency dependent.
Under this choice, a $1\%$ change in amplitude width and a $1~\mathrm{deg}$ change in phase width are treated on a comparable footing.
Other choices may be appropriate for different calibration goals, and the dependence on this ratio is examined separately in Fig.~\ref{fig:score_color} and \ref{app:468}.

For each analysis segment \(\mathcal{T}_j\), the near-optimal evaluation frequency is defined as the frequency that minimizes this score over the candidate set \(\mathcal{F}_{\mathrm{cand}}\):
\begin{equation}
f_{e,\mathrm{opt}}^{(j)}
=
\arg\min_{f_e \in \mathcal{F}_{\mathrm{cand}}}
S^{(j)}(f_e,\sigma_{|A|},\sigma_\phi).
\label{eq:fe_opt}
\end{equation}
For visualization, we also use the normalized score
\begin{equation}
S_{\mathrm{norm}}^{(j)}(f_e,\sigma_{|A|},\sigma_\phi)
=
\frac{
S^{(j)}(f_e,\sigma_{|A|},\sigma_\phi)
}{
\min_{f_e' \in \mathcal{F}_{\mathrm{cand}}}
S^{(j)}(f_e',\sigma_{|A|},\sigma_\phi)
},
\label{eq:score_norm}
\end{equation}
which is normalized so that the minimum value within each segment is unity.
To quantify how clearly the best candidate is separated from the second-best one, we further define
\begin{equation}
\Delta S_{2\mathrm{nd}}^{(j)}
=
\frac{
S^{(j)}(f_{e,2\mathrm{nd}}^{(j)},\sigma_{|A|},\sigma_\phi)
-
S^{(j)}(f_{e,\mathrm{opt}}^{(j)},\sigma_{|A|},\sigma_\phi)
}{
S^{(j)}(f_{e,\mathrm{opt}}^{(j)},\sigma_{|A|},\sigma_\phi)
}.
\label{eq:score_2nd}
\end{equation}
Here, \(f_{e,2\mathrm{nd}}^{(j)}\) denotes the second-best candidate in the segment \(\mathcal{T}_j\).

The score defined here is intended to rank candidate evaluation frequencies only from the viewpoint of statistical uncertainty, and does not include the bias or uncertainty introduced by the frequency translation.

\subsection{Analysis setup for KAGRA O4a}

We apply the framework defined in Secs.~\ref{sec:sens} and \ref{sec:obj} to KAGRA O4a data.
The procedure used to estimate \(\kappa_C(t)\) and \(f_{\mathrm{cp}}(t)\) from the calibration lines follows \cite{Hido2026}. In this subsection, we summarize the analysis setup used to compare the sensing function and the score across evaluation frequencies.

The candidate evaluation-frequency set \(\mathcal{F}_{\mathrm{cand}}\) is defined from the frequency points used in the transfer-function measurements during O4a. These candidates are common to all analysis segments. However, frequency points below 20~Hz are excluded from the candidate set, because the influence of the intermediate-mass stage of the suspension is no longer negligible in that range~\cite{IM_info}.

For the score evaluation, we adopt the frequency band from \(30\) to \(1500~\mathrm{Hz}\). The quantities defined in Sec.~\ref{sec:obj} are evaluated on a logarithmically spaced grid of 1000 frequency points over this band, and the score is computed by numerical integration.

The analysis uses the full O4a dataset. Since the score is defined segment by segment, the data are divided into analysis segments \(\mathcal{T}_j\) of duration \(4096~\mathrm{s}\). Segments shorter than 4096~s are not used. For continuous stretches longer than 4096~s, as many 4096~s analysis segments as possible are extracted, with the remaining time distributed as evenly as possible between the two ends, so that the segmentation is centered within each continuous stretch.

\section{Statistical optimization results}
\label{sec:results}

\subsection{Representative segment}

\begin{figure}
 \centering
    \includegraphics[width=0.7\textwidth]{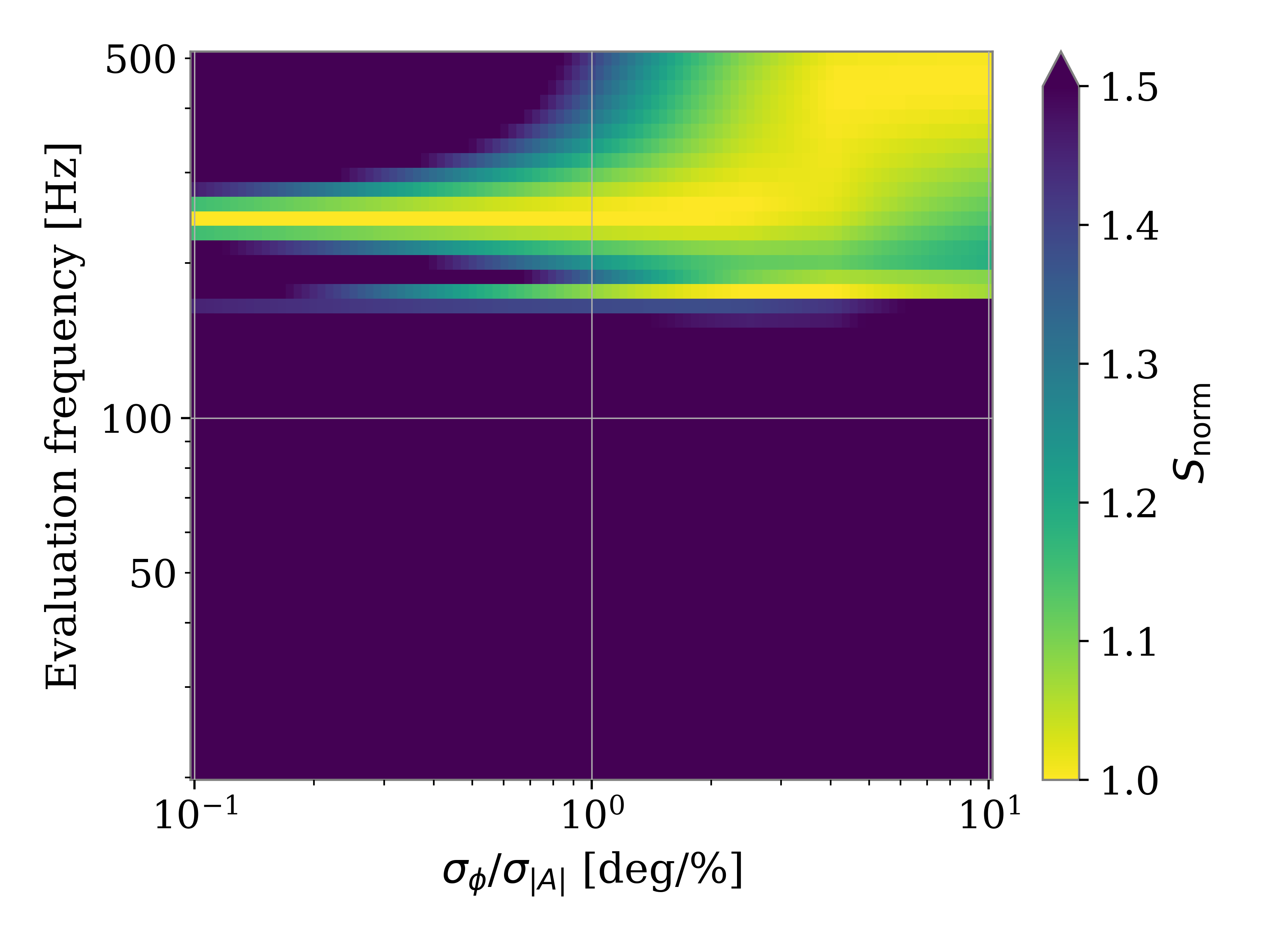}
 \caption{Representative heatmap of the normalized score $S_{\mathrm{norm}}$
for a 4096~s O4a segment, plotted as a function of the evaluation frequency $f_e$
and the normalization ratio $\sigma_\phi/\sigma_{|A|}$.
A broad minimum appears near $244$~Hz for $\sigma_\phi/\sigma_{|A|} \lesssim 2$,
and the preferred frequency shifts upward as the magnitude term is weighted more strongly.
}
 \label{fig:score_color}
\end{figure}

We first examine a representative 4096~s segment. Fig.~\ref{fig:score_color} shows the normalized score \(S_{\mathrm{norm}}\) as a function of the evaluation frequency \(f_e\) and the normalization ratio \(\sigma_{\phi}/\sigma_{|A|}\).
Fig.~\ref{fig:score_color} shows a broad minimum near 244~Hz over a wide range of \(\sigma_{\phi}/\sigma_{|A|}\). In particular, for \(\sigma_{\phi}/\sigma_{|A|} \lesssim 2\), the minimum is obtained at 244~Hz. As \(\sigma_{\phi}/\sigma_{|A|}\) increases, the preferred evaluation frequency shifts toward higher frequencies, indicating that the near-optimal frequency depends on the relative weighting of the magnitude and phase terms in the score.

In the following, we adopt \(\sigma_{\phi}/\sigma_{|A|}=1\) as the representative setting. Under this choice, 244~Hz is selected as the best candidate in the representative segment. Although higher frequencies become more competitive when the magnitude term is weighted more strongly, 244~Hz remains preferred under the combined objective adopted here. \ref{app:468} illustrates this point by comparison with the 468~Hz candidate, which yields a smaller amplitude interval width over much of the band but at the cost of a substantially larger phase interval width.

\subsection{O4a-wide summary of the optimal candidate}

\begin{figure}
 \centering
    \includegraphics[width=0.6\textwidth]{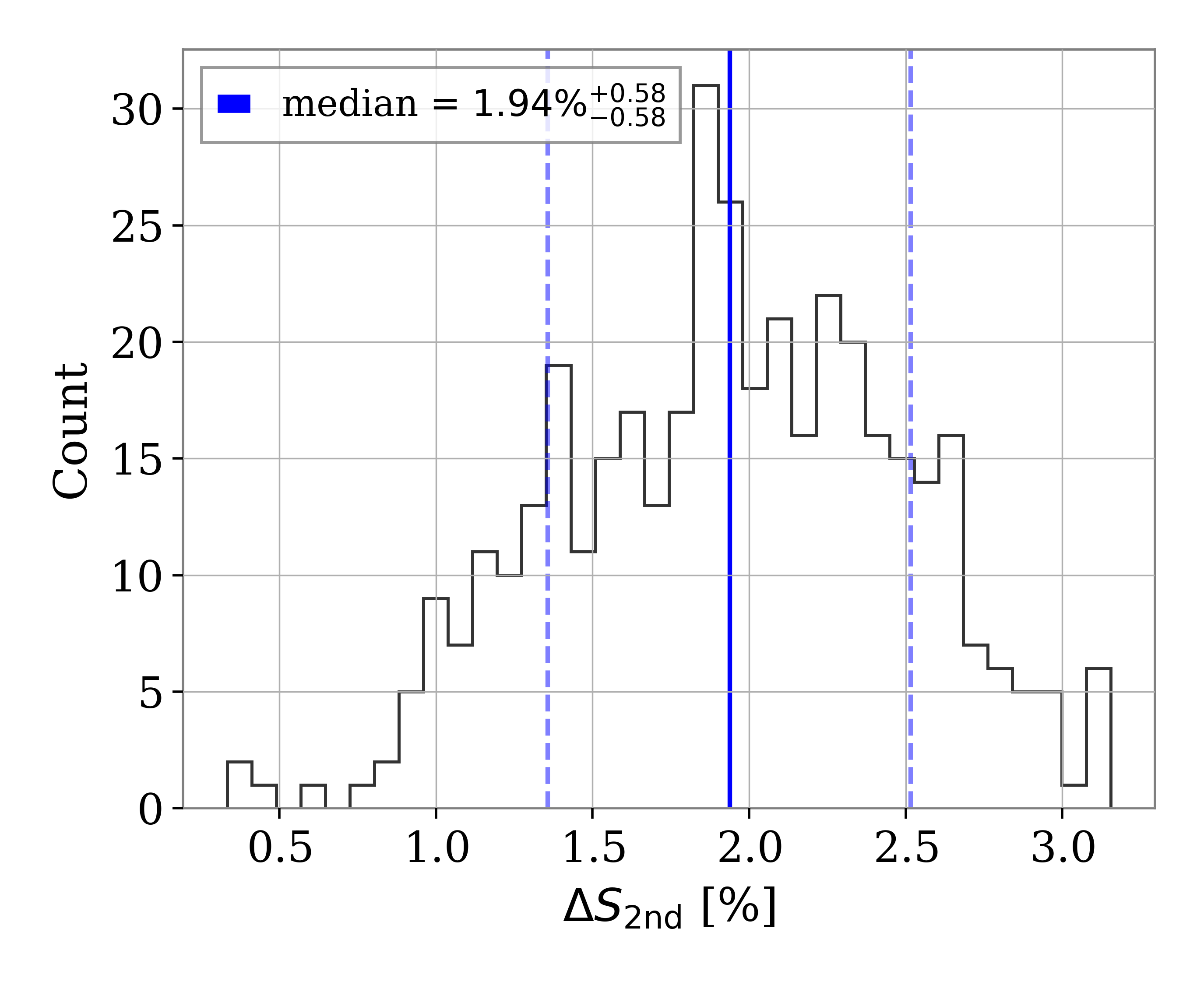}
 \caption{Histogram of the relative score difference between the best candidate at
244~Hz and the second-best candidate over all 4096~s segments in the KAGRA O4a data,
for the representative normalization ratio $\sigma_\phi/\sigma_{|A|}=1$.
For each segment, the quantity $\Delta S_{2\mathrm{nd}}$ is evaluated.
The solid vertical line indicates the median, and the dashed vertical lines indicate
the 16th and 84th percentiles.
The distribution shows that 244~Hz is the best candidate throughout O4a and yields
a typically $\sim 2\%$ smaller score than the second-best candidate.
}
 \label{fig:score_2nd}
\end{figure}

Next, we assess whether the trend seen in the representative segment persists throughout the full O4a period. For each 4096~s segment, we evaluate the relative score difference between the best candidate and the second-best candidate. Fig.~\ref{fig:score_2nd} summarizes the distribution of \(\Delta S_{2\mathrm{nd}}\), defined in Sec.~\ref{sec:obj}, over all analysis segments in O4a for \(\sigma_{\phi}/\sigma_{|A|}=1\).

Across O4a, 244~Hz is selected as the best candidate in all 4096~s segments, with a typical score advantage of about \(2\%\) over the second-best candidate. We therefore use 244~Hz as the representative best candidate in the following.

\subsection{Reduction of statistical uncertainty}
\label{sec:reduc}

\begin{figure}
 \centering
    \includegraphics[width=0.65\textwidth]{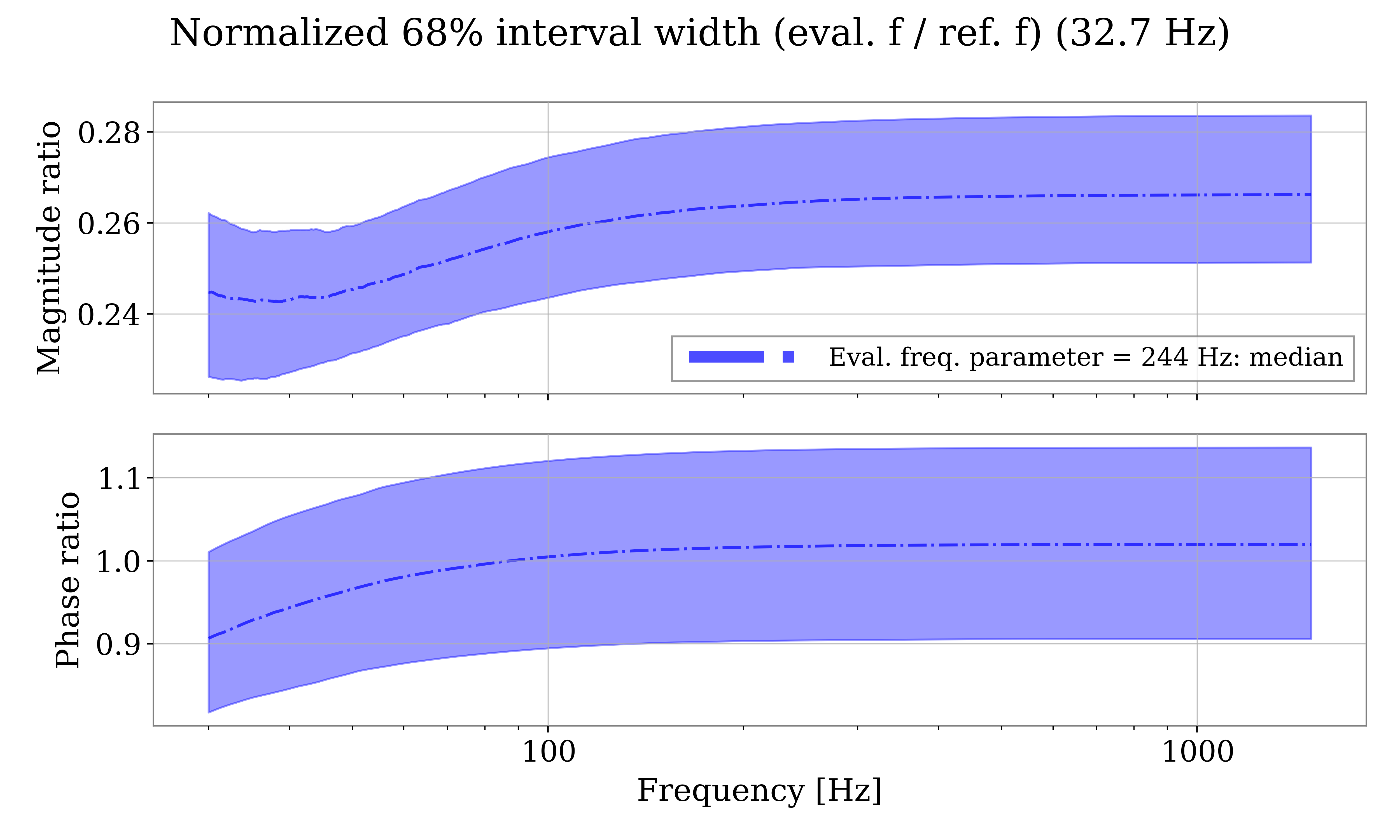}
 \caption{
O4a-wide summary of the \(68\%\) interval widths of the sample distribution
for the amplitude and phase of the sensing function, evaluated at \(244~\mathrm{Hz}\)
for all \(4096~\mathrm{s}\) segments.
The horizontal axis denotes the frequency \(f\).
At each frequency, the median and the 16th--84th percentiles over segments are shown
relative to the reference-frequency result at \(32.7~\mathrm{Hz}\).
The phase interval width remains broadly unchanged, while the amplitude interval width
is markedly reduced over a broad frequency range, reaching about one quarter of the reference value.
 }
 \label{fig:cdeg_ui}
\end{figure}

Fig.~\ref{fig:cdeg_ui} summarizes the O4a-wide change in the 68\% interval widths of the amplitude and phase of the sensing function when the evaluation frequency is changed from the reference value of 32.7~Hz to 244~Hz.

Using 244~Hz leaves the phase interval width broadly unchanged while substantially reducing the amplitude interval width over a broad frequency range, typically to about one quarter of the reference result at 32.7~Hz. This indicates that, under the statistical objective adopted here, the main advantage of the 244~Hz evaluation frequency appears as a marked improvement in the amplitude-side statistical uncertainty.

\section{Supplementary validation of translation-induced discrepancy}
\label{sec:supp}

\begin{figure}
 \centering
    \includegraphics[width=0.8\textwidth]{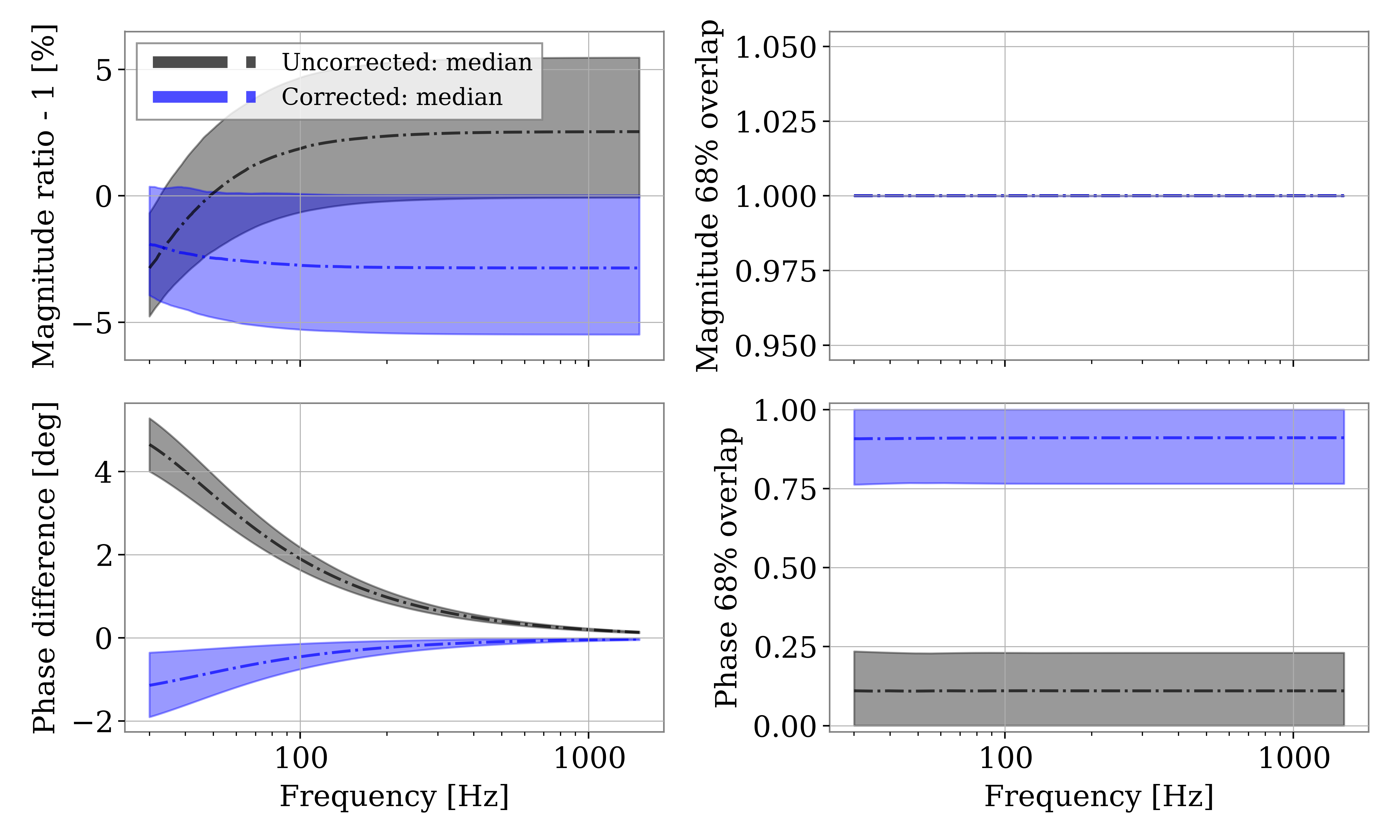}
 \caption{
O4a-wide summary of the discrepancy between the reference-frequency result at
\(32.7~\mathrm{Hz}\) and the translated result at \(244~\mathrm{Hz}\).
The left panel shows the bias estimated from the difference between the segment-wise
medians, and the right panel shows the overlap ratio of the corresponding \(68\%\) intervals
of the sample distribution, defined as the overlap length normalized by the smaller interval width.
At each frequency, the median and the 16th--84th percentiles over all \(4096~\mathrm{s}\) segments
are shown for the uncorrected and model-measurement-bias-corrected cases.
The uncorrected translation introduces a clear phase bias and poorer interval overlap,
whereas the model-measurement bias correction reduces these discrepancies, especially in phase.
}
 \label{fig:bias_all}
\end{figure}

We now assess the discrepancy introduced by frequency translation by comparing the reference-frequency result at 32.7~Hz with the translated result at 244~Hz.

Fig.~\ref{fig:bias_all} summarizes this discrepancy over all 4096~s segments in O4a. The left panel shows the bias estimated from the difference between the segment-wise medians, while the right panel shows the overlap ratio of the corresponding 68\% intervals. Here, the overlap ratio is defined as the overlap length normalized by the smaller of the two interval widths. To examine the effect of the model--measurement bias correction, we show both the uncorrected and corrected cases. The correction procedure follows Ref.~\cite{Hido2026}, and the uncertainty contribution of the correction factor itself is not included in this comparison.

In the uncorrected case, the discrepancy is most evident in phase, showing both a non-negligible median-level bias and poorer interval overlap relative to the reference-frequency result. In amplitude, a frequency-dependent median-level difference is also visible, although the corresponding 68\% intervals remain more broadly consistent.

Applying the model--measurement bias correction reduces this discrepancy substantially, especially in phase. The median-level phase bias becomes smaller, and the interval overlap improves over a broad frequency range. In amplitude, the frequency dependence seen in the uncorrected case is mitigated, although a residual median-level bias remains.


\section{Discussion}
\label{sec:discussion}

This study identifies near-optimal evaluation frequencies only with respect to the statistical criterion introduced here.
For KAGRA O4a, the representative choice \(\sigma_{\phi}/\sigma_{|A|}=1\) selects 244~Hz as the preferred candidate, substantially reducing the amplitude-side statistical uncertainty while leaving the phase interval width broadly comparable to that at 32.7~Hz.
At the same time, the comparison in Sec.~\ref{sec:supp} shows that frequency translation can introduce a non-negligible discrepancy when model--measurement bias is present, most clearly in phase.
These results should therefore be interpreted as a supplementary validation of translation consistency.
The uncertainty contribution of the bias-correction factor itself remains outside the present uncertainty budget.

Accordingly, the practical role of the proposed method is to preselect statistically promising evaluation frequencies under operational constraints.
In practice, calibration-line frequencies are usually fixed before the observing run begins, so the present framework is most naturally used to identify promising candidates in advance and then screen them further using independent information on model--measurement bias.
Frequencies that require little or no additional bias correction are especially attractive in such an application, because they reduce the dependence on translation-induced discrepancy and additional uncertainty.
More generally, the dependence of the preferred frequency on the relative weighting between amplitude and phase suggests that the framework can be adapted to different calibration goals.
The same idea may also be useful in other calibration schemes that rely on calibration lines and reference-model-based translation under operational constraints.

From this practical perspective, it is worth asking why the statistical criterion used here favors an intermediate evaluation frequency.
A simple error-propagation argument gives some qualitative intuition.
In the sensing-side calculation, part of the procedure involves differences between complex transfer-function terms~\cite{Tuyenbayev_2017}, and the propagated absolute uncertainty is given by the quadrature sum of the contributing uncertainties.
As a result, the relative effect of the propagated uncertainty can be smaller when the resulting difference is larger in magnitude.
However, because the sensing-side quantities are obtained through a nonlinear transformation of this complex quantity, this intuition alone does not determine the preferred evaluation frequency.
We therefore rely on the empirical score defined in Sec.~\ref{sec:obj} for the quantitative ranking.

\section{Conclusion}
\label{sec:conclusion}

We presented a statistical framework for preselecting evaluation frequencies for sensing-side calibration without changing the injected calibration-line frequency itself. Applied to KAGRA O4a, the framework identifies 244~Hz as the preferred candidate under the representative choice \(\sigma_{\phi}/\sigma_{|A|}=1\), substantially reducing the amplitude-side statistical uncertainty while leaving the phase interval width broadly comparable to that at 32.7~Hz.

This result should be interpreted as a statistical preselection outcome rather than as a final ranking under a full uncertainty budget, since the uncertainty contribution associated with translation-related bias correction is not included in the present analysis.
Within this scope, the proposed method provides an effective statistical framework for preselecting candidate frequencies under operational constraints before an observing run.

\FloatBarrier
\section{Acknowledgments}

This work was supported by MEXT, JSPS Leading-edge Research Infrastructure Program, JSPS Grant-in-Aid for Specially Promoted Research 26000005, JSPS Grant-in-Aid for Scientific Research on Innovative Areas 2402: 24103006, 24103005, and 2905: JP17H06358, JP17H06361 and JP17H06364, JSPS Core-to-Core Program A. Advanced Research Networks, JSPS Grant-in-Aid for Scientific Research (S) 17H06133 and 20H05639 , JSPS Grant-in-Aid for Transformative Research Areas (A) 20A203: JP20H05854, the joint research program of the Institute for Cosmic Ray Research, University of Tokyo, National Research Foundation (NRF), Computing Infrastructure Project of Global Science experimental Data hub Center (GSDC) at KISTI, Korea Astronomy and Space Science Institute (KASI), and Ministry of Science and ICT (MSIT) in Korea, Academia Sinica (AS), AS Grid Center (ASGC) and the National Science and Technology Council (NSTC) in Taiwan under grants including the Science Vanguard Research Program, Advanced Technology Center (ATC) of NAOJ, and Mechanical Engineering Center of KEK.

\appendix
\renewcommand{\thesection}{Appendix \Alph{section}}
\section{Representative comparison of the sensing function}
\label{app:repC}

\begin{figure}
 \centering
 \begin{subfigure}[b]{0.495\linewidth}
    \centering
    \includegraphics[width=0.99\textwidth]{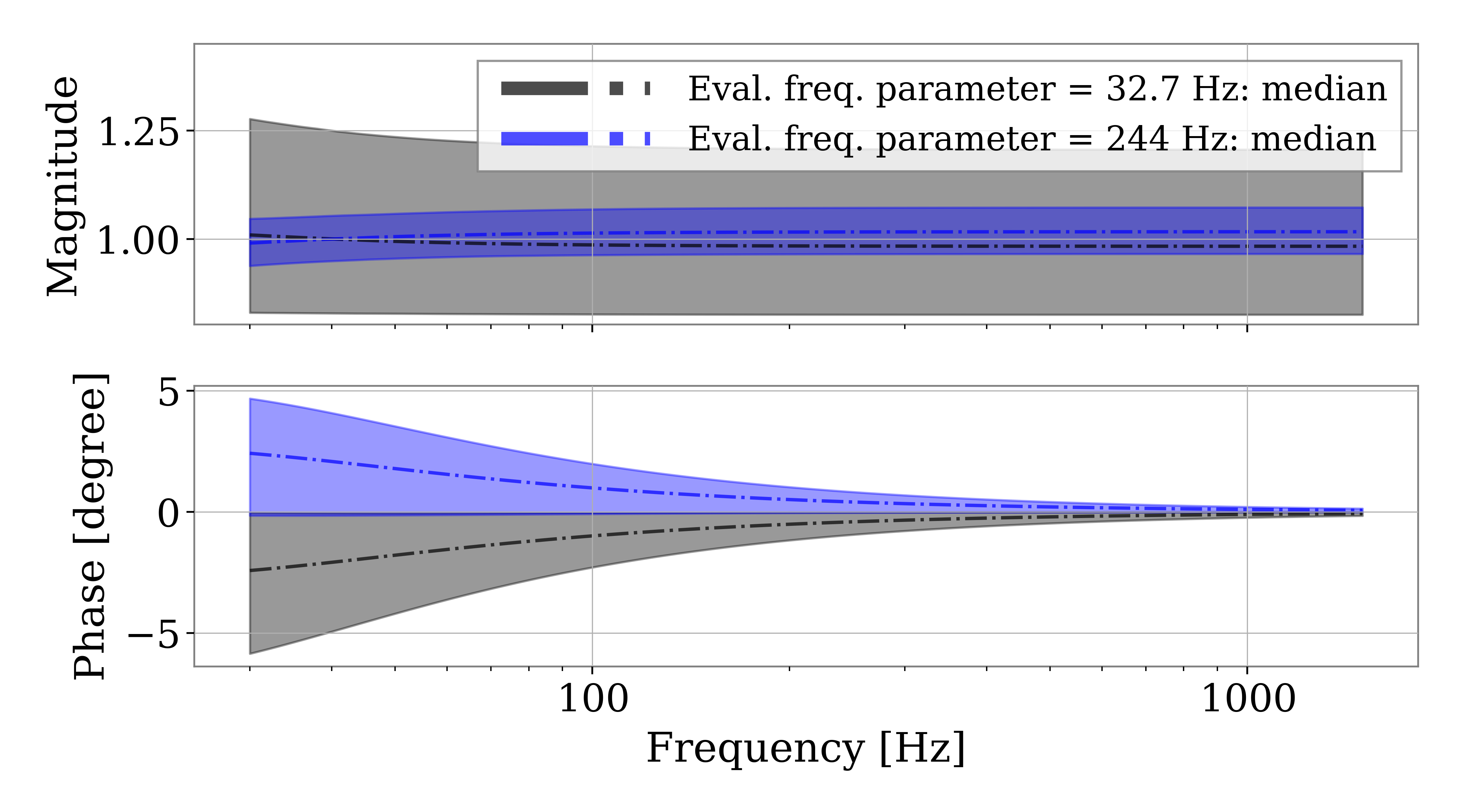}
     \caption{Uncorrected}
    \label{fig:c_a}
 \end{subfigure}
  \hfill
 \begin{subfigure}[b]{0.495\linewidth}
    \centering
    \includegraphics[width=0.99\textwidth]{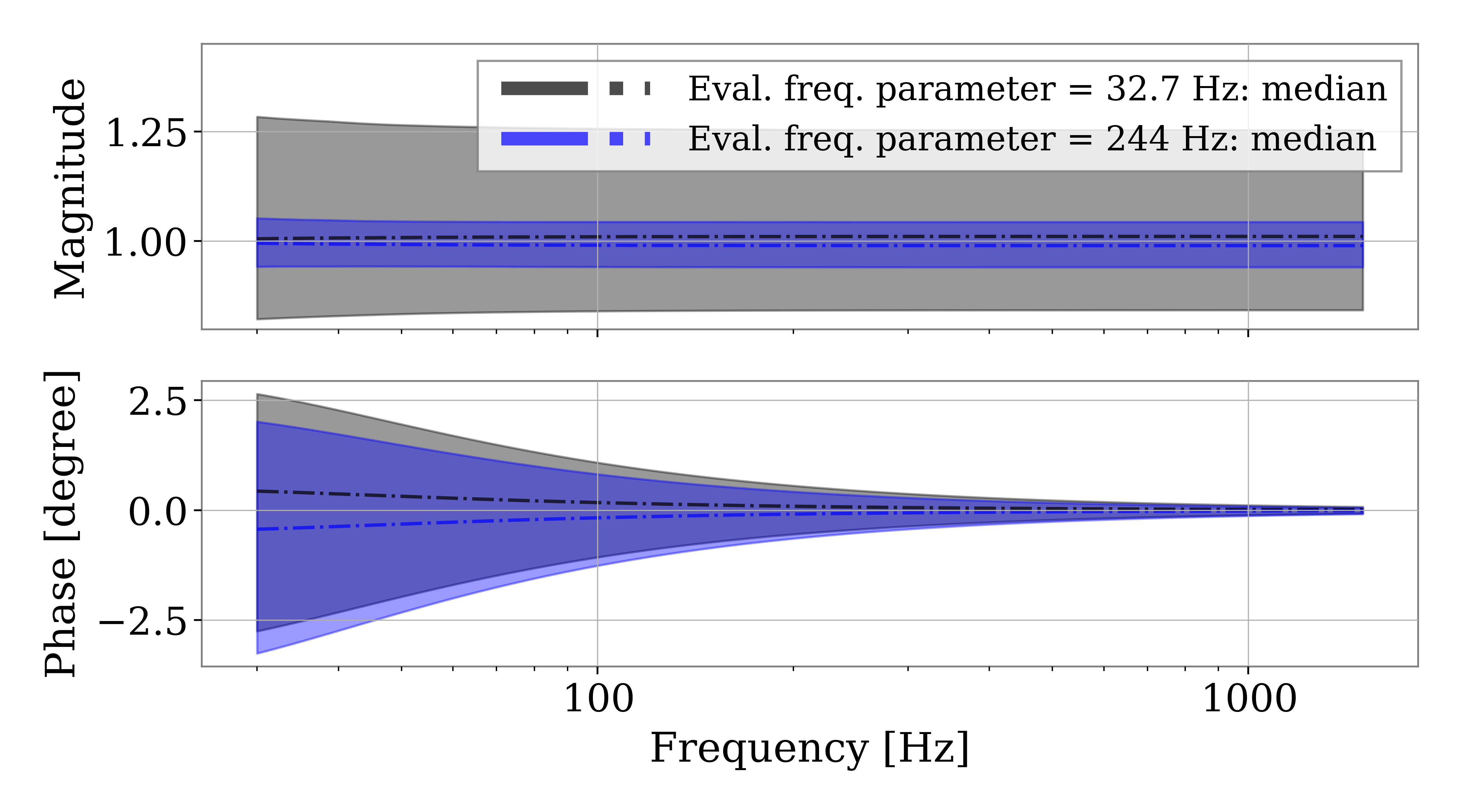}
     \caption{Corrected}
    \label{fig:c_frh}
 \end{subfigure}
 \caption{
Representative comparison of the sensing function in a single \(4096~\mathrm{s}\) O4a segment.
The upper and lower panels show the amplitude and phase, respectively,
for the reference-frequency result at \(32.7~\mathrm{Hz}\) and the translated result at \(244~\mathrm{Hz}\).
Panels (a) and (b) show the uncorrected and model-measurement-bias-corrected cases.
The median and the corresponding \(68\%\) intervals of the sample distribution are shown.
The translated result at \(244~\mathrm{Hz}\) exhibits a smaller 68\% amplitude interval width,
while the uncorrected comparison shows a visible discrepancy, particularly in phase.
This discrepancy is reduced after applying the model-measurement bias correction.
The uncertainty contribution of the correction factor itself is not included.
 }
 \label{fig:c_all}
\end{figure}

To illustrate the results of Secs.~\ref{sec:reduc} and \ref{sec:supp} more directly,
Fig.~\ref{fig:c_all} shows a representative \(4096~\mathrm{s}\) segment from O4a,
comparing the sensing function obtained at the reference frequency of \(32.7~\mathrm{Hz}\)
with that obtained by translation to \(244~\mathrm{Hz}\).
The upper and lower panels show the amplitude and phase, respectively,
and panels \ref{fig:c_a} and \ref{fig:c_frh} correspond to the uncorrected and corrected cases.

This representative example makes the trends discussed in the main text visually clear.
The translated result at \(244~\mathrm{Hz}\) retains a narrower \(68\%\) interval in amplitude
than the reference-frequency result, consistent with the reduction in statistical uncertainty
summarized in Sec.~\ref{sec:reduc}.
At the same time, in the uncorrected comparison, a clear discrepancy is visible,
most notably in phase.
After applying the model-measurement bias correction,
this discrepancy is reduced, especially in phase,
while the smaller amplitude interval at \(244~\mathrm{Hz}\) is largely retained.

\section{Comparison with the 468~Hz candidate}
\label{app:468}

\begin{figure}
 \centering
    \includegraphics[width=0.7\textwidth]{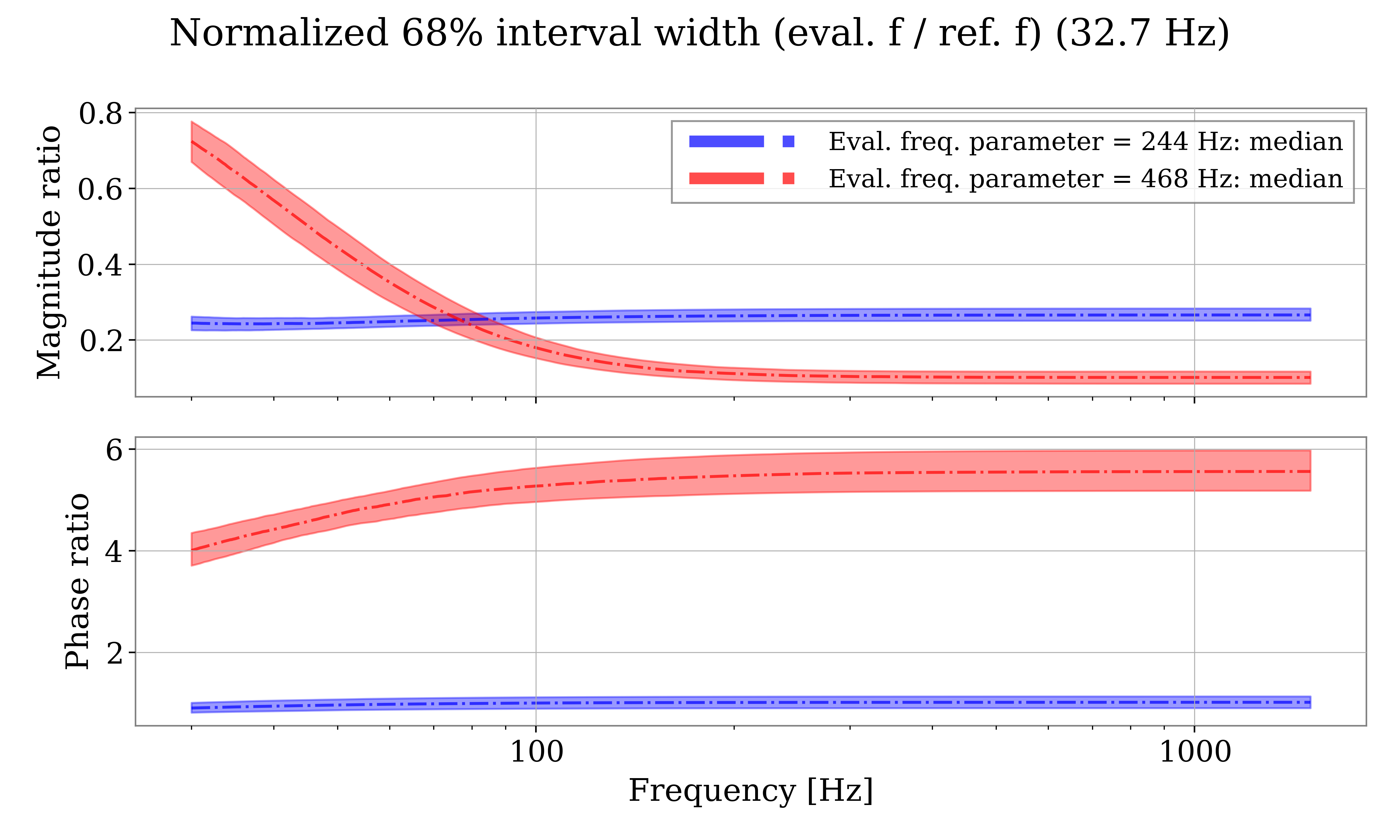}
 \caption{
Comparison of the normalized \(68\%\) interval widths of the statistical uncertainty
for the sensing function, evaluated at \(244~\mathrm{Hz}\) and \(468~\mathrm{Hz}\)
and shown relative to the reference-frequency result at \(32.7~\mathrm{Hz}\).
The upper and lower panels show the amplitude and phase results, respectively.
While \(468~\mathrm{Hz}\) yields a smaller amplitude interval width than \(244~\mathrm{Hz}\)
over much of the frequency range, it also leads to a substantially larger phase interval width.
This comparison illustrates why \(468~\mathrm{Hz}\), although more favorable in amplitude,
is less favorable than \(244~\mathrm{Hz}\) under the combined objective adopted in this work.
}
 \label{fig:cdeg_468}
\end{figure}

As suggested by the heatmap in Fig.~\ref{fig:score_color}, the \(468~\mathrm{Hz}\) candidate is competitive when the magnitude term is weighted more strongly. To illustrate this point more directly, Fig.~\ref{fig:cdeg_468} compares the normalized \(68\%\) interval widths of the sensing function for the \(244~\mathrm{Hz}\) and \(468~\mathrm{Hz}\) candidates, both shown relative to the reference-frequency result at \(32.7~\mathrm{Hz}\).

The figure shows that the \(468~\mathrm{Hz}\) candidate yields a smaller amplitude interval width than \(244~\mathrm{Hz}\) over much of the frequency range. In this sense, \(468~\mathrm{Hz}\) is more favorable than \(244~\mathrm{Hz}\) for amplitude alone. On the other hand, the phase interval width at \(468~\mathrm{Hz}\) is markedly larger than that at \(244~\mathrm{Hz}\) across the full band considered.

This comparison clarifies why \(468~\mathrm{Hz}\), although advantageous in amplitude, is not selected as the preferred candidate under the combined objective adopted in this work. The result is consistent with Fig.~\ref{fig:score_color}, where higher evaluation frequencies become more favorable only when the magnitude term is given relatively larger weight.

\printbibliography

@article{Tuyenbayev_2017,
doi = {10.1088/0264-9381/34/1/015002},
url = {https://doi.org/10.1088/0264-9381/34/1/015002},
year = {2016},
month = {dec},
publisher = {IOP Publishing},
volume = {34},
number = {1},
pages = {015002},
author = {Tuyenbayev, D and Karki, S and Betzwieser, J and Cahillane, C and Goetz, E and Izumi, K and Kandhasamy, S and Kissel, J S and Mendell, G and Wade, M and Weinstein, A J and Savage, R L},
title = {Improving LIGO calibration accuracy by tracking and compensating for slow temporal variations},
journal = {Classical and Quantum Gravity}
}

@article{PhysRevD.85.064034,
  title = {Effect of calibration errors on Bayesian parameter estimation for gravitational wave signals from inspiral binary systems in the advanced detectors era},
  author = {Vitale, Salvatore and Del Pozzo, Walter and Li, Tjonnie G. F. and Van Den Broeck, Chris and Mandel, Ilya and Aylott, Ben and Veitch, John},
  journal = {Phys. Rev. D},
  volume = {85},
  issue = {6},
  pages = {064034},
  numpages = {18},
  year = {2012},
  month = {Mar},
  publisher = {American Physical Society},
  doi = {10.1103/PhysRevD.85.064034},
  url = {https://link.aps.org/doi/10.1103/PhysRevD.85.064034}
}

@article{PhysRevD.111.063034,
  title = {Impact of calibration uncertainties on Hubble constant measurements from gravitational-wave sources},
  author = {Huang, Yiwen and Chen, Hsin-Yu and Haster, Carl-Johan and Sun, Ling and Vitale, Salvatore and Kissel, Jeffrey S.},
  journal = {Phys. Rev. D},
  volume = {111},
  issue = {6},
  pages = {063034},
  numpages = {12},
  year = {2025},
  month = {Mar},
  publisher = {American Physical Society},
  doi = {10.1103/PhysRevD.111.063034},
  url = {https://link.aps.org/doi/10.1103/PhysRevD.111.063034}
}

@article{Hall_2019,
doi = {10.1088/1361-6382/ab368c},
url = {https://doi.org/10.1088/1361-6382/ab368c},
year = {2019},
month = {sep},
publisher = {IOP Publishing},
volume = {36},
number = {20},
pages = {205006},
author = {Hall, Evan D and Cahillane, Craig and Izumi, Kiwamu and Smith, Rory J E and Adhikari, Rana X},
title = {Systematic calibration error requirements for gravitational-wave detectors via the Cramér–Rao bound},
journal = {Classical and Quantum Gravity}
}

@article{10.1093/ptep/ptaa125,
    author = {Akutsu, T and Ando, M and Arai, K and Arai, Y and Araki, S and Araya, A and Aritomi, N and Aso, Y and Bae, S and Bae, Y and Baiotti, L and Bajpai, R and Barton, M A and Cannon, K and Capocasa, E and Chan, M and Chen, C and Chen, K and Chen, Y and Chu, H and Chu, Y -K and Eguchi, S and Enomoto, Y and Flaminio, R and Fujii, Y and Fukunaga, M and Fukushima, M and Ge, G and Hagiwara, A and Haino, S and Hasegawa, K and Hayakawa, H and Hayama, K and Himemoto, Y and Hiranuma, Y and Hirata, N and Hirose, E and Hong, Z and Hsieh, B H and Huang, C -Z and Huang, P and Huang, Y and Ikenoue, B and Imam, S and Inayoshi, K and Inoue, Y and Ioka, K and Itoh, Y and Izumi, K and Jung, K and Jung, P and Kajita, T and Kamiizumi, M and Kanda, N and Kang, G and Kawaguchi, K and Kawai, N and Kawasaki, T and Kim, C and Kim, J C and Kim, W S and Kim, Y -M and Kimura, N and Kita, N and Kitazawa, H and Kojima, Y and Kokeyama, K and Komori, K and Kong, A K H and Kotake, K and Kozakai, C and Kozu, R and Kumar, R and Kume, J and Kuo, C and Kuo, H -S and Kuroyanagi, S and Kusayanagi, K and Kwak, K and Lee, H K and Lee, H W and Lee, R and Leonardi, M and Lin, L C -C and Lin, C -Y and Lin, F -L and Liu, G C and Luo, L -W and Marchio, M and Michimura, Y and Mio, N and Miyakawa, O and Miyamoto, A and Miyazaki, Y and Miyo, K and Miyoki, S and Morisaki, S and Moriwaki, Y and Nagano, K and Nagano, S and Nakamura, K and Nakano, H and Nakano, M and Nakashima, R and Narikawa, T and Negishi, R and Ni, W -T and Nishizawa, A and Obuchi, Y and Ogaki, W and Oh, J J and Oh, S H and Ohashi, M and Ohishi, N and Ohkawa, M and Okutomi, K and Oohara, K and Ooi, C P and Oshino, S and Pan, K and Pang, H and Park, J and Arellano, F E Peña and Pinto, I and Sago, N and Saito, S and Saito, Y and Sakai, K and Sakai, Y and Sakuno, Y and Sato, S and Sato, T and Sawada, T and Sekiguchi, T and Sekiguchi, Y and Shibagaki, S and Shimizu, R and Shimoda, T and Shimode, K and Shinkai, H and Shishido, T and Shoda, A and Somiya, K and Son, E J and Sotani, H and Sugimoto, R and Suzuki, T and Suzuki, T and Tagoshi, H and Takahashi, H and Takahashi, R and Takamori, A and Takano, S and Takeda, H and Takeda, M and Tanaka, H and Tanaka, K and Tanaka, K and Tanaka, T and Tanaka, T and Tanioka, S and Tapia San Martin, E N and Telada, S and Tomaru, T and Tomigami, Y and Tomura, T and Travasso, F and Trozzo, L and Tsang, T and Tsubono, K and Tsuchida, S and Tsuzuki, T and Tuyenbayev, D and Uchikata, N and Uchiyama, T and Ueda, A and Uehara, T and Ueno, K and Ueshima, G and Uraguchi, F and Ushiba, T and van Putten, M H P M and Vocca, H and Wang, J and Wu, C and Wu, H and Wu, S and Xu, W- R and Yamada, T and Yamamoto, K and Yamamoto, K and Yamamoto, T and Yokogawa, K and Yokoyama, J and Yokozawa, T and Yoshioka, T and Yuzurihara, H and Zeidler, S and Zhao, Y and Zhu, Z -H},
    title = {Overview of KAGRA: Detector design and construction history},
    journal = {Progress of Theoretical and Experimental Physics},
    volume = {2021},
    number = {5},
    pages = {05A101},
    year = {2020},
    month = {08},
    issn = {2050-3911},
    doi = {10.1093/ptep/ptaa125},
    url = {https://doi.org/10.1093/ptep/ptaa125},
    eprint = {https://academic.oup.com/ptep/article-pdf/2021/5/05A101/37974994/ptaa125.pdf},
}

@article{Aasi_2015,
doi = {10.1088/0264-9381/32/7/074001},
url = {https://doi.org/10.1088/0264-9381/32/7/074001},
year = {2015},
month = {mar},
publisher = {IOP Publishing},
volume = {32},
number = {7},
pages = {074001},
author = {The LIGO Scientific Collaboration and Aasi, J and Abbott, B P and Abbott, R and Abbott, T and Abernathy, M R and Ackley, K and Adams, C and Adams, T and Addesso, P and Adhikari, R X and Adya, V and Affeldt, C and Aggarwal, N and Aguiar, O D and Ain, A and Ajith, P and Alemic, A and Allen, B and Amariutei, D and Anderson, S B and Anderson, W G and Arai, K and Araya, M C and Arceneaux, C and Areeda, J S and Ashton, G and Ast, S and Aston, S M and Aufmuth, P and Aulbert, C and Aylott, B E and Babak, S and Baker, P T and Ballmer, S W and Barayoga, J C and Barbet, M and Barclay, S and Barish, B C and Barker, D and Barr, B and Barsotti, L and Bartlett, J and Barton, M A and Bartos, I and Bassiri, R and Batch, J C and Baune, C and Behnke, B and Bell, A S and Bell, C and Benacquista, M and Bergman, J and Bergmann, G and Berry, C P L and Betzwieser, J and Bhagwat, S and Bhandare, R and Bilenko, I A and Billingsley, G and Birch, J and Biscans, S and Biwer, C and Blackburn, J K and Blackburn, L and Blair, C D and Blair, D and Bock, O and Bodiya, T P and Bojtos, P and Bond, C and Bork, R and Born, M and Bose, Sukanta and Brady, P R and Braginsky, V B and Brau, J E and Bridges, D O and Brinkmann, M and Brooks, A F and Brown, D A and Brown, D D and Brown, N M and Buchman, S and Buikema, A and Buonanno, A and Cadonati, L and Calderón Bustillo, J and Camp, J B and Cannon, K C and Cao, J and Capano, C D and Caride, S and Caudill, S and Cavaglià, M and Cepeda, C and Chakraborty, R and Chalermsongsak, T and Chamberlin, S J and Chao, S and Charlton, P and Chen, Y and Cho, H S and Cho, M and Chow, J H and Christensen, N and Chu, Q and Chung, S and Ciani, G and Clara, F and Clark, J A and Collette, C and Cominsky, L and Constancio, M and Cook, D and Corbitt, T R and Cornish, N and Corsi, A and Costa, C A and Coughlin, M W and Countryman, S and Couvares, P and Coward, D M and Cowart, M J and Coyne, D C and Coyne, R and Craig, K and Creighton, J D E and Creighton, T D and Cripe, J and Crowder, S G and Cumming, A and Cunningham, L and Cutler, C and Dahl, K and Dal Canton, T and Damjanic, M and Danilishin, S L and Danzmann, K and Dartez, L and Dave, I and Daveloza, H and Davies, G S and Daw, E J and DeBra, D and Del Pozzo, W and Denker, T and Dent, T and Dergachev, V and DeRosa, R T and DeSalvo, R and Dhurandhar, S and D´ıaz, M and Di Palma, I and Dojcinoski, G and Dominguez, E and Donovan, F and Dooley, K L and Doravari, S and Douglas, R and Downes, T P and Driggers, J C and Du, Z and Dwyer, S and Eberle, T and Edo, T and Edwards, M and Edwards, M and Effler, A and Eggenstein, H.-B and Ehrens, P and Eichholz, J and Eikenberry, S S and Essick, R and Etzel, T and Evans, M and Evans, T and Factourovich, M and Fairhurst, S and Fan, X and Fang, Q and Farr, B and Farr, W M and Favata, M and Fays, M and Fehrmann, H and Fejer, M M and Feldbaum, D and Ferreira, E C and Fisher, R P and Frei, Z and Freise, A and Frey, R and Fricke, T T and Fritschel, P and Frolov, V V and Fuentes-Tapia, S and Fulda, P and Fyffe, M and Gair, J R and Gaonkar, S and Gehrels, N and Gergely´, L Á and Giaime, J A and Giardina, K D and Gleason, J and Goetz, E and Goetz, R and Gondan, L and González, G and Gordon, N and Gorodetsky, M L and Gossan, S and Goßler, S and Gräf, C and Graff, P B and Grant, A and Gras, S and Gray, C and Greenhalgh, R J S and Gretarsson, A M and Grote, H and Grunewald, S and Guido, C J and Guo, X and Gushwa, K and Gustafson, E K and Gustafson, R and Hacker, J and Hall, E D and Hammond, G and Hanke, M and Hanks, J and Hanna, C and Hannam, M D and Hanson, J and Hardwick, T and Harry, G M and Harry, I W and Hart, M and Hartman, M T and Haster, C-J and Haughian, K and Hee, S and Heintze, M and Heinzel, G and Hendry, M and Heng, I S and Heptonstall, A W and Heurs, M and Hewitson, M and Hild, S and Hoak, D and Hodge, K A and Hollitt, S E and Holt, K and Hopkins, P and Hosken, D J and Hough, J and Houston, E and Howell, E J and Hu, Y M and Huerta, E and Hughey, B and Husa, S and Huttner, S H and Huynh, M and Huynh-Dinh, T and Idrisy, A and Indik, N and Ingram, D R and Inta, R and Islas, G and Isler, J C and Isogai, T and Iyer, B R and Izumi, K and Jacobson, M and Jang, H and Jawahar, S and Ji, Y and Jiménez-Forteza, F and Johnson, W W and Jones, D I and Jones, R and Ju, L and Haris, K and Kalogera, V and Kandhasamy, S and Kang, G and Kanner, J B and Katsavounidis, E and Katzman, W and Kaufer, H and Kaufer, S and Kaur, T and Kawabe, K and Kawazoe, F and Keiser, G M and Keitel, D and Kelley, D B and Kells, W and Keppel, D G and Key, J S and Khalaidovski, A and Khalili, F Y and Khazanov, E A and Kim, C and Kim, K and Kim, N G and Kim, N and Kim, Y.-M and King, E J and King, P J and Kinzel, D L and Kissel, J S and Klimenko, S and Kline, J and Koehlenbeck, S and Kokeyama, K and Kondrashov, V and Korobko, M and Korth, W Z and Kozak, D B and Kringel, V and Krishnan, B and Krueger, C and Kuehn, G and Kumar, A and Kumar, P and Kuo, L and Landry, M and Lantz, B and Larson, S and Lasky, P D and Lazzarini, A and Lazzaro, C and Le, J and Leaci, P and Leavey, S and Lebigot, E O and Lee, C H and Lee, H K and Lee, H M and Leong, J R and Levin, Y and Levine, B and Lewis, J and Li, T G F and Libbrecht, K and Libson, A and Lin, A C and Littenberg, T B and Lockerbie, N A and Lockett, V and Logue, J and Lombardi, A L and Lormand, M and Lough, J and Lubinski, M J and Lück, H and Lundgren, A P and Lynch, R and Ma, Y and Macarthur, J and MacDonald, T and Machenschalk, B and MacInnis, M and Macleod, D M and Magaña-Sandoval, F and Magee, R and Mageswaran, M and Maglione, C and Mailand, K and Mandel, I and Mandic, V and Mangano, V and Mansell, G L and Márka, S and Márka, Z and Markosyan, A and Maros, E and Martin, I W and Martin, R M and Martynov, D and Marx, J N and Mason, K and Massinger, T J and Matichard, F and Matone, L and Mavalvala, N and Mazumder, N and Mazzolo, G and McCarthy, R and McClelland, D E and McCormick, S and McGuire, S C and McIntyre, G and McIver, J and McLin, K and McWilliams, S and Meadors, G D and Meinders, M and Melatos, A and Mendell, G and Mercer, R A and Meshkov, S and Messenger, C and Meyers, P M and Miao, H and Middleton, H and Mikhailov, E E and Miller, A and Miller, J and Millhouse, M and Ming, J and Mirshekari, S and Mishra, C and Mitra, S and Mitrofanov, V P and Mitselmakher, G and Mittleman, R and Moe, B and Mohanty, S D and Mohapatra, S R P and Moore, B and Moraru, D and Moreno, G and Morriss, S R and Mossavi, K and Mow-Lowry, C M and Mueller, C L and Mueller, G and Mukherjee, S and Mullavey, A and Munch, J and Murphy, D and Murray, P G and Mytidis, A and Nash, T and Nayak, R K and Necula, V and Nedkova, K and Newton, G and Nguyen, T and Nielsen, A B and Nissanke, S and Nitz, A H and Nolting, D and Normandin, M E N and Nuttall, L K and Ochsner, E and O’Dell, J and Oelker, E and Ogin, G H and Oh, J J and Oh, S H and Ohme, F and Oppermann, P and Oram, R and O’Reilly, B and Ortega, W and O’Shaughnessy, R and Osthelder, C and Ott, C D and Ottaway, D J and Ottens, R S and Overmier, H and Owen, B J and Padilla, C and Pai, A and Pai, S and Palashov, O and Pal-Singh, A and Pan, H and Pankow, C and Pannarale, F and Pant, B C and Papa, M A and Paris, H and Patrick, Z and Pedraza, M and Pekowsky, L and Pele, A and Penn, S and Perreca, A and Phelps, M and Pierro, V and Pinto, I M and Pitkin, M and Poeld, J and Post, A and Poteomkin, A and Powell, J and Prasad, J and Predoi, V and Premachandra, S and Prestegard, T and Price, L R and Principe, M and Privitera, S and Prix, R and Prokhorov, L and Puncken, O and Pürrer, M and Qin, J and Quetschke, V and Quintero, E and Quiroga, G and Quitzow-James, R and Raab, F J and Rabeling, D S and Radkins, H and Raffai, P and Raja, S and Rajalakshmi, G and Rakhmanov, M and Ramirez, K and Raymond, V and Reed, C M and Reid, S and Reitze, D H and Reula, O and Riles, K and Robertson, N A and Robie, R and Rollins, J G and Roma, V and Romano, J D and Romanov, G and Romie, J H and Rowan, S and Rüdiger, A and Ryan, K and Sachdev, S and Sadecki, T and Sadeghian, L and Saleem, M and Salemi, F and Sammut, L and Sandberg, V and Sanders, J R and Sannibale, V and Santiago-Prieto, I and Sathyaprakash, B S and Saulson, P R and Savage, R and Sawadsky, A and Scheuer, J and Schilling, R and Schmidt, P and Schnabel, R and Schofield, R M S and Schreiber, E and Schuette, D and Schutz, B F and Scott, J and Scott, S M and Sellers, D and Sengupta, A S and Sergeev, A and Serna, G and Sevigny, A and Shaddock, D A and Shahriar, M S and Shaltev, M and Shao, Z and Shapiro, B and Shawhan, P and Shoemaker, D H and Sidery, T L and Siemens, X and Sigg, D and Silva, A D and Simakov, D and Singer, A and Singer, L and Singh, R and Sintes, A M and Slagmolen, B J J and Smith, J R and Smith, M R and Smith, R J E and Smith-Lefebvre, N D and Son, E J and Sorazu, B and Souradeep, T and Staley, A and Stebbins, J and Steinke, M and Steinlechner, J and Steinlechner, S and Steinmeyer, D and Stephens, B C and Steplewski, S and Stevenson, S and Stone, R and Strain, K A and Strigin, S and Sturani, R and Stuver, A L and Summerscales, T Z and Sutton, P J and Szczepanczyk, M and Szeifert, G and Talukder, D and Tanner, D B and Tápai, M and Tarabrin, S P and Taracchini, A and Taylor, R and Tellez, G and Theeg, T and Thirugnanasambandam, M P and Thomas, M and Thomas, P and Thorne, K A and Thorne, K S and Thrane, E and Tiwari, V and Tomlinson, C and Torres, C V and Torrie, C I and Traylor, G and Tse, M and Tshilumba, D and Ugolini, D and Unnikrishnan, C S and Urban, A L and Usman, S A and Vahlbruch, H and Vajente, G and Valdes, G and Vallisneri, M and van Veggel, A A and Vass, S and Vaulin, R and Vecchio, A and Veitch, J and Veitch, P J and Venkateswara, K and Vincent-Finley, R and Vitale, S and Vo, T and Vorvick, C and Vousden, W D and Vyatchanin, S P and Wade, A R and Wade, L and Wade, M and Walker, M and Wallace, L and Walsh, S and Wang, H and Wang, M and Wang, X and Ward, R L and Warner, J and Was, M and Weaver, B and Weinert, M and Weinstein, A J and Weiss, R and Welborn, T and Wen, L and Wessels, P and Westphal, T and Wette, K and Whelan, J T and Whitcomb, S E and White, D J and Whiting, B F and Wilkinson, C and Williams, L and Williams, R and Williamson, A R and Willis, J L and Willke, B and Wimmer, M and Winkler, W and Wipf, C C and Wittel, H and Woan, G and Worden, J and Xie, S and Yablon, J and Yakushin, I and Yam, W and Yamamoto, H and Yancey, C C and Yang, Q and Zanolin, M and Zhang, Fan and Zhang, L and Zhang, M and Zhang, Y and Zhao, C and Zhou, M and Zhu, X J and Zucker, M E and Zuraw, S and Zweizig, J},
title = {Advanced LIGO},
journal = {Classical and Quantum Gravity}
}

@article{Acernese_2015,
doi = {10.1088/0264-9381/32/2/024001},
url = {https://doi.org/10.1088/0264-9381/32/2/024001},
year = {2014},
month = {dec},
publisher = {IOP Publishing},
volume = {32},
number = {2},
pages = {024001},
author = {Acernese, F and Agathos, M and Agatsuma, K and Aisa, D and Allemandou, N and Allocca, A and Amarni, J and Astone, P and Balestri, G and Ballardin, G and Barone, F and Baronick, J-P and Barsuglia, M and Basti, A and Basti, F and Bauer, Th S and Bavigadda, V and Bejger, M and Beker, M G and Belczynski, C and Bersanetti, D and Bertolini, A and Bitossi, M and Bizouard, M A and Bloemen, S and Blom, M and Boer, M and Bogaert, G and Bondi, D and Bondu, F and Bonelli, L and Bonnand, R and Boschi, V and Bosi, L and Bouedo, T and Bradaschia, C and Branchesi, M and Briant, T and Brillet, A and Brisson, V and Bulik, T and Bulten, H J and Buskulic, D and Buy, C and Cagnoli, G and Calloni, E and Campeggi, C and Canuel, B and Carbognani, F and Cavalier, F and Cavalieri, R and Cella, G and Cesarini, E and Mottin, E Chassande- and Chincarini, A and Chiummo, A and Chua, S and Cleva, F and Coccia, E and Cohadon, P-F and Colla, A and Colombini, M and Conte, A and Coulon, J-P and Cuoco, E and Dalmaz, A and D’Antonio, S and Dattilo, V and Davier, M and Day, R and Debreczeni, G and Degallaix, J and Deléglise, S and Pozzo, W Del and Dereli, H and Rosa, R De and Fiore, L Di and Lieto, A Di and Virgilio, A Di and Doets, M and Dolique, V and Drago, M and Ducrot, M and Endrőczi, G and Fafone, V and Farinon, S and Ferrante, I and Ferrini, F and Fidecaro, F and Fiori, I and Flaminio, R and Fournier, J-D and Franco, S and Frasca, S and Frasconi, F and Gammaitoni, L and Garufi, F and Gaspard, M and Gatto, A and Gemme, G and Gendre, B and Genin, E and Gennai, A and Ghosh, S and Giacobone, L and Giazotto, A and Gouaty, R and Granata, M and Greco, G and Groot, P and Guidi, G M and Harms, J and Heidmann, A and Heitmann, H and Hello, P and Hemming, G and Hennes, E and Hofman, D and Jaranowski, P and Jonker, R J G and Kasprzack, M and Kéfélian, F and Kowalska, I and Kraan, M and Królak, A and Kutynia, A and Lazzaro, C and Leonardi, M and Leroy, N and Letendre, N and Li, T G F and Lieunard, B and Lorenzini, M and Loriette, V and Losurdo, G and Magazzù, C and Majorana, E and Maksimovic, I and Malvezzi, V and Man, N and Mangano, V and Mantovani, M and Marchesoni, F and Marion, F and Marque, J and Martelli, F and Martellini, L and Masserot, A and Meacher, D and Meidam, J and Mezzani, F and Michel, C and Milano, L and Minenkov, Y and Moggi, A and Mohan, M and Montani, M and Morgado, N and Mours, B and Mul, F and Nagy, M F and Nardecchia, I and Naticchioni, L and Nelemans, G and Neri, I and Neri, M and Nocera, F and Pacaud, E and Palomba, C and Paoletti, F and Paoli, A and Pasqualetti, A and Passaquieti, R and Passuello, D and Perciballi, M and Petit, S and Pichot, M and Piergiovanni, F and Pillant, G and Piluso, A and Pinard, L and Poggiani, R and Prijatelj, M and Prodi, G A and Punturo, M and Puppo, P and Rabeling, D S and Rácz, I and Rapagnani, P and Razzano, M and Re, V and Regimbau, T and Ricci, F and Robinet, F and Rocchi, A and Rolland, L and Romano, R and Rosińska, D and Ruggi, P and Saracco, E and Sassolas, B and Schimmel, F and Sentenac, D and Sequino, V and Shah, S and Siellez, K and Straniero, N and Swinkels, B and Tacca, M and Tonelli, M and Travasso, F and Turconi, M and Vajente, G and van Bakel, N and van Beuzekom, M and van den Brand, J F J and Van Den Broeck, C and van der Sluys, M V and van Heijningen, J and Vasúth, M and Vedovato, G and Veitch, J and Verkindt, D and Vetrano, F and Viceré, A and Vinet, J-Y and Visser, G and Vocca, H and Ward, R and Was, M and Wei, L-W and Yvert, M and żny, A Zadro and Zendri, J-P},
title = {Advanced Virgo: a second-generation interferometric gravitational wave detector},
journal = {Classical and Quantum Gravity}
}

@article{Wade_2023,
doi = {10.1088/1361-6382/acabf6},
url = {https://doi.org/10.1088/1361-6382/acabf6},
year = {2023},
month = {jan},
publisher = {IOP Publishing},
volume = {40},
number = {3},
pages = {035001},
author = {Wade, M and Viets, A D and Chmiel, T and Stover, M and Wade, L},
title = {Improving LIGO calibration accuracy by using time-dependent filters to compensate for temporal variations},
journal = {Classical and Quantum Gravity}
}

@article{Viets_2018,
doi = {10.1088/1361-6382/aab658},
url = {https://doi.org/10.1088/1361-6382/aab658},
year = {2018},
month = {apr},
publisher = {IOP Publishing},
volume = {35},
number = {9},
pages = {095015},
author = {Viets, A D and Wade, M and Urban, A L and Kandhasamy, S and Betzwieser, J and Brown, Duncan A and Burguet-Castell, J and Cahillane, C and Goetz, E and Izumi, K and Karki, S and Kissel, J S and Mendell, G and Savage, R L and Siemens, X and Tuyenbayev, D and Weinstein, A J},
title = {Reconstructing the calibrated strain signal in the Advanced LIGO detectors},
journal = {Classical and Quantum Gravity}
}

@misc{IM_info,
  author = {Hido Shingo and Yamamoto Takahiro},
  title = {MN stage in the offline/low-latency reconstruction pipelines},
  year = {2024},
  url = {https://klog.icrr.u-tokyo.ac.jp/osl/?r=29691},
  note = {(Accessed: 2026-04-01)}
}

@misc{LINE_info,
  author = {Yamamoto Takahiro},
  title = {Candidate of new frequencies for calibration lines},
  year = {2023},
  url = {https://klog.icrr.u-tokyo.ac.jp/osl/?r=25235},
  note = {(Accessed: 2026-04-27)}
}

@article{10.1093/ptep/ptab018,
    author = {Akutsu, T and Ando, M and Arai, K and Arai, Y and Araki, S and Araya, A and Aritomi, N and Asada, H and Aso, Y and Bae, S and Bae, Y and Baiotti, L and Bajpai, R and Barton, M A and Cannon, K and Cao, Z and Capocasa, E and Chan, M and Chen, C and Chen, K and Chen, Y and Chiang, C -Y and Chu, H and Chu, Y -K and Eguchi, S and Enomoto, Y and Flaminio, R and Fujii, Y and Fujikawa, Y and Fukunaga, M and Fukushima, M and Gao, D and Ge, G and Ha, S and Hagiwara, A and Haino, S and Han, W -B and Hasegawa, K and Hattori, K and Hayakawa, H and Hayama, K and Himemoto, Y and Hiranuma, Y and Hirata, N and Hirose, E and Hong, Z and Hsieh, B and Huang, G -Z and Huang, H -Y and Huang, P and Huang, Y -C and Huang, Y and Hui, D  C  Y and Ide, S and Ikenoue, B and Imam, S and Inayoshi, K and Inoue, Y and Ioka, K and Ito, K and Itoh, Y and Izumi, K and Jeon, C and Jin, H -B and Jung, K and Jung, P and Kaihotsu, K and Kajita, T and Kakizaki, M and Kamiizumi, M and Kanda, N and Kang, G and Kawaguchi, K and Kawai, N and Kawasaki, T and Kim, C and Kim, J and Kim, J C and Kim, W S and Kim, Y -M and Kimura, N and Kita, N and Kitazawa, H and Kojima, Y and Kokeyama, K and Komori, K and Kong, A K H and Kotake, K and Kozakai, C and Kozu, R and Kumar, R and Kume, J and Kuo, C and Kuo, H -S and Kuromiya, Y and Kuroyanagi, S and Kusayanagi, K and Kwak, K and Lee, H K and Lee, H W and Lee, R and Leonardi, M and Li, K L and Lin, L C -C and Lin, C -Y and Lin, F -K and Lin, F -L and Lin, H L and Liu, G C and Luo, L -W and Majorana, E and Marchio, M and Michimura, Y and Mio, N and Miyakawa, O and Miyamoto, A and Miyazaki, Y and Miyo, K and Miyoki, S and Mori, Y and Morisaki, S and Moriwaki, Y and Nagano, K and Nagano, S and Nakamura, K and Nakano, H and Nakano, M and Nakashima, R and Nakayama, Y and Narikawa, T and Naticchioni, L and Negishi, R and Nguyen Quynh, L and Ni, W -T and Nishizawa, A and Nozaki, S and Obuchi, Y and Ogaki, W and Oh, J J and Oh, K and Oh, S H and Ohashi, M and Ohishi, N and Ohkawa, M and Ohta, H and Okutani, Y and Okutomi, K and Oohara, K and Ooi, C and Oshino, S and Otabe, S and Pan, K and Pang, H and Parisi, A and Park, J and Peña Arellano, F E and Pinto, I and Sago, N and Saito, S and Saito, Y and Sakai, K and Sakai, Y and Sakuno, Y and Sato, S and Sato, T and Sawada, T and Sekiguchi, T and Sekiguchi, Y and Shao, L and Shibagaki, S and Shimizu, R and Shimoda, T and Shimode, K and Shinkai, H and Shishido, T and Shoda, A and Somiya, K and Son, E J and Sotani, H and Sugimoto, R and Suresh, J and Suzuki, T and Suzuki, T and Tagoshi, H and Takahashi, H and Takahashi, R and Takamori, A and Takano, S and Takeda, H and Takeda, M and Tanaka, H and Tanaka, K and Tanaka, K and Tanaka, T and Tanaka, T and Tanioka, S and Tapia San Martin, E N and Telada, S and Tomaru, T and Tomigami, Y and Tomura, T and Travasso, F and Trozzo, L and Tsang, T and Tsao, J -S and Tsubono, K and Tsuchida, S and Tsutsui, T and Tsuzuki, T and Tuyenbayev, D and Uchikata, N and Uchiyama, T and Ueda, A and Uehara, T and Ueno, K and Ueshima, G and Uraguchi, F and Ushiba, T and van Putten, M H P M and Vocca, H and Wang, J and Washimi, T and Wu, C and Wu, H and Wu, S and Xu, W -R and Yamada, T and Yamamoto, K and Yamamoto, K and Yamamoto, T and Yamashita, K and Yamazaki, R and Yang, Y and Yokogawa, K and Yokoyama, J and Yokozawa, T and Yoshioka, T and Yuzurihara, H and Zeidler, S and Zhan, M and Zhang, H and Zhao, Y and Zhu, Z -H},
    title = {Overview of KAGRA: Calibration, detector characterization, physical environmental monitors, and the geophysics interferometer},
    journal = {Progress of Theoretical and Experimental Physics},
    volume = {2021},
    number = {5},
    pages = {05A102},
    year = {2021},
    month = {02},
    issn = {2050-3911},
    doi = {10.1093/ptep/ptab018},
    url = {https://doi.org/10.1093/ptep/ptab018},
    eprint = {https://academic.oup.com/ptep/article-pdf/2021/5/05A102/38109702/ptab018.pdf},
}

@misc{Hido2026,
      title={Statistical Estimation and Correction of Model-Measurement Bias in Time-Dependent Correction Factors of KAGRA}, 
      author={Shingo Hido and Takahiro Yamamoto and Dan Chen and Takahiro Sawada and Shinji Miyoki},
      year={2026},
      eprint={2606.09010},
      archivePrefix={arXiv},
      primaryClass={astro-ph.IM},
      url={https://arxiv.org/abs/2606.09010}, 
}

\end{document}